\documentclass[twocolumn,showpacs,preprintnumbers,amsmath,amssymb]{revtex4}
\usepackage{graphicx}
\usepackage{dcolumn}
\usepackage{bm}

\newcommand{\hH}{\hat{H}}
\newcommand{\hW}{\hat{W}}
\newcommand{\ha}{\hat{a}}
\newcommand{\hap}{\hat{a}^+}

\begin{document}

\title{Theory of electron Zitterbewegung in graphene probed by femtosecond laser pulses}
\date{\today}
\author{Tomasz M. Rusin*}
\email{Tomasz.Rusin@centertel.pl}
\author{Wlodek Zawadzki\dag}
\affiliation{*PTK Centertel Sp. z o.o., ul. Skierniewicka 10A, 01-230 Warsaw, Poland\\
            \dag Institute of Physics, Polish Academy of Sciences, Al. Lotnik\'ow 32/46, 02-688 Warsaw, Poland}

\begin{abstract}
We propose an experiment allowing an observation of Zitterbewegung (ZB, trembling motion) of
electrons in graphene in the presence of a magnetic field.
In contrast to the existing theoretical work we make no assumptions concerning shape of the
electron wave packet. A femtosecond Gaussian laser pulse excites electrons from the
valence $n=-1$ Landau level into three other levels,
creating an oscillating electron wave packet with interband and intraband frequencies.
Oscillations of an average position of the
packet are directly related to the induced dipole moment and oscillations
of the average packet's acceleration determine emitted electric field. Both quantities
can be measured experimentally. A broadening of Landau levels is included to make the description
of ZB as realistic as possible. Criteria of realization of a ZB experiment are discussed.
\end{abstract}

\maketitle
\section{Introduction}

Zitterbewegung (ZB, trembling motion) of relativistic electrons in a vacuum was
predicted nearly 80 years ago by Schrodinger \cite{Schroedinger30}.
Unfortunately, both the spacial extension of the ZB motion, being of the order of
$\lambda_c=\hbar/mc$, and the ZB frequency $\omega_Z=2mc^2/\hbar$ are far beyond
current experimental possibilities.
However, it was recently shown that, because of an analogy between the behavior of relativistic
electrons in a vacuum and that of electrons in narrow gap semiconductors
\cite{ZawadzkiOPS,ZawadzkiHMF}, one can expect
the trembling motion of electrons in narrow gap semiconductors having much more advantageous
characteristics: the frequency $\omega_Z=E_g/\hbar$ and the amplitude $\lambda_Z=\hbar/m_0^*u$,
where $E_g$ is the energy gap, $m_0^*$ is the electron effective mass, and $u=(E_g/2m_0^*)^{1/2}$
is the maximum electron velocity in the two-band energy spectrum \cite{Zawadzki05KP}.
It was further shown that the ZB-like
motion should occur in other two-band situations, both in solids
\cite{Cannata90,Schliemann05,Zawadzki06,Katsnelson06,Winkler07,Auslender07,
Rusin07a,Rusin07b,Trauzettel07,Zulick07,Rusin08a,Zawadzki08,Englman08,Demikhovskii08}
and in other systems \cite{Cserti06,Lamata07,Clark07,Zhang08a,Merkl08,Bermudez08,Martinez08,Dora09}.
An observation of an acoustic analogue of ZB was reported recently in sonic crystals \cite{Zhang08b}.

It was predicted some time ago by Lock \cite{Lock79} that,
if an electron is represented by a wave packet, the
ZB phenomenon will have a transient character. This was confirmed by very recent calculations
which predicted that the decay times of ZB are of the order of femtoseconds to microseconds depending
on the system in question \cite{Rusin07b,Zawadzki08,Clark07}.
However, it was also shown that the presence of an
external magnetic field and the resulting Landau quantization of the electron spectrum 'stabilizes'
the situation making the ZB oscillations stationary in time, if one neglects the loss of
electron energy due to dipole radiation \cite{Rusin08a, Schliemann08}.
It is known that an external magnetic field does not induce interband electron transitions,
so that an interference of electron states corresponding to positive and negative energies
remains unchanged and an appearance of interband frequencies remains the signature of ZB phenomenon.
On the other hand, due to Landau quantization of the electron and hole energies also
intraband (cyclotron) excitations appear in the spectrum.

All the recent theoretical work on ZB assumed that initially the electrons are represented by
Gaussian wave packets \cite{Huang52,Schliemann05,Rusin07b,Rusin08a,Clark07,Zhang08a,Merkl08,Bermudez08,Lock79}.
While this assumption represents a real progress compared to the initial work
that had treated electrons as plain waves
\cite{Cannata90,Zawadzki05KP,Zawadzki06,Katsnelson06,Winkler07,Rusin07a,Cserti06,GreinerBook,BjorkenBook},
it is obviously an idealization since it is
not quite clear how to prepare an electron in this form. It is the purpose of our present work
to propose and describe an observation of electron ZB in semiconductors that can be really carried out.
Namely, we calculate a reaction of an electron in graphene excited by a laser pulse,
not assuming anything about initial form of the electron wave packet.
In our description we take into account currently available experimental possibilities.
Also, we include a broadening of Landau levels and investigate its effect on the
trembling motion.
It is our hope that this proposition will help to observe for the first time this somewhat
mysterious effect that is fundamental for both relativistic electrons in a vacuum and electrons in
narrow gap semiconductors.

The following conditions should be met for a successful observation of ZB:
a) The ZB frequency must be in the range of currently detectable regimes, i.e.
 of the order of  $\omega_Z \approx $ 1 fs$^{-1}$,   and the
   size of oscillations should be of the order of a few \AA;
b) The ZB oscillations should be persistent or slowly transient;
c) Both positive and negative electron energies must be excited with a sufficient probability;
d) To avoid many-electron effects (see Refs. \cite{Krekora2004,Wang08}),  the wave packet
   should be created in a one-electron regime.
A system that in our opinion fulfills the above criteria is p-type monolayer graphene
in a constant magnetic field. The wave packet should be created by an ultra short monocycle or
sub-monocycle laser pulse. Because of a very wide frequency spectrum of such a pulse, the resulting
wave packet will have both positive and negative energies. The electron oscillations
give rise to a time-dependent dipole moment which will be a source of
electric field and it will emit or absorb radiation in the far infrared range.
Experimental parameters necessary to create the optical wave packet and to detect the radiation
should be within the current experimental possibilities.

Our paper is organized as follows.  In sections II and III we calculate the electron reaction to a
short laser pulse, the creation of ZB, and we describe an electric field of radiation emitted by the
trembling electron. In Section IV the influence of Landau level broadening on the
trembling motion is investigated. In Section V we describe the time-dependent luminescence
filtered by a time gate and a frequency filter.
Finally, we discuss our results and conclude by a summary.

\section{Preliminaries}
\begin{figure}
\includegraphics[width=8.5cm,height=8.cm]{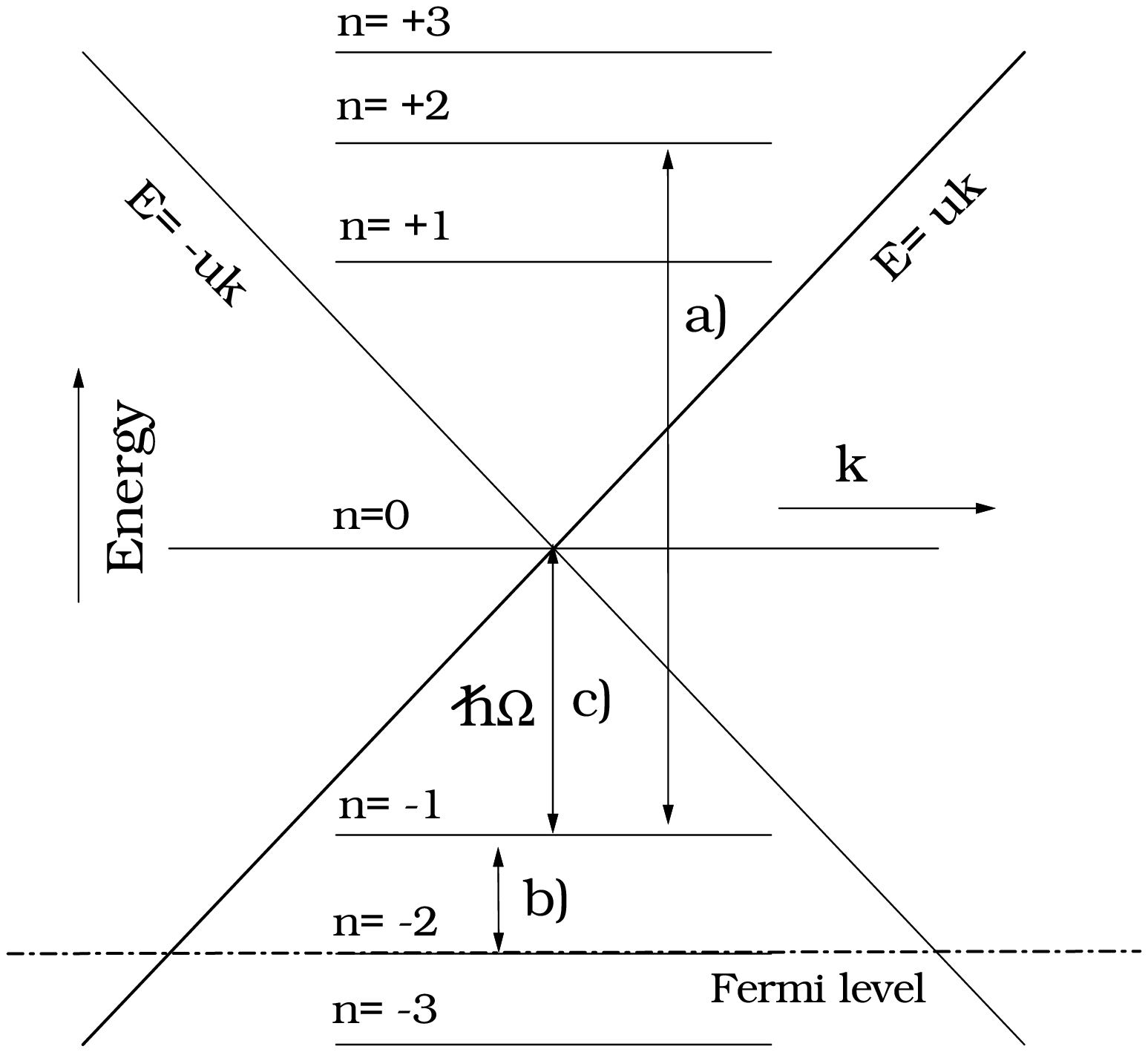}
\caption{Electron energy levels for monolayer graphene in a magnetic field.
 Proposed position of the Fermi level is indicated. Arrows show interband (a), intraband (b),
 and fundamental (c) energies, see text.} \label{Fig0}
\end{figure}

We consider p-type monolayer graphene in the presence of a  magnetic field perpendicular to the layer.
The electron-hole Hamiltonian $\hH_1$ at the $K_1$ point of the Brillouin zone is given by
\cite{Novoselov04,Novoselov05,Gusynin07}
\begin{equation}  \label{H aap}
 \hH_1 = -\hbar\Omega\left(\begin{array}{cc}  0 & \ha \\ \hap& 0 \\  \end{array}\right),
\end{equation}
where $\ha$, $\hap$ are the lowering and rising operators and the characteristic
frequency of the system is $\Omega=\sqrt{2}u/L$, where $L=\sqrt{\hbar/eB}$ is the magnetic length,
and the velocity $u\simeq$ 10$^{8}$ cm/s.
The energy spectrum of $\hH_1$ is
$E_{n}={\rm sgn}(n)\hbar\Omega\sqrt{|n|}$, where $n=0,\pm 1,\ldots$, see Fig. \ref{Fig0}.
The eigenstates of Hamiltonian (\ref{H aap}) for the gauge $\bm A = [-By,0,0]$ are
\begin{equation} \label{H_psi}
\psi_n(\bm r)= \frac{e^{ik_xx}}{\sqrt{4\pi}}
               \left(\begin{array}{c} -{\rm sgn}(n)\phi_{|n|-1}(\xi) \\ \phi_{|n|}(\xi)    \end{array}\right),
\end{equation}
where $\xi=y/L-k_xL$ and $\phi_n(\xi)$=$(C_n/\sqrt{L})\ e^{-\frac{1}{2}\xi^2}{\rm H}_n(\xi)$
is the $n$-th eigenstate of the harmonic oscillator. Here
$C_n=1/\sqrt{2^nn!\sqrt{\pi}}\ $, and ${\rm H}_n(\xi)$ are the Hermite polynomials.
For $n=0$, the first component in Eq. (\ref{H_psi}) vanishes and the normalization coefficient
is $1/\sqrt{2\pi}$. We assume the Fermi level to coincide with the Landau level (LL) $n=-2$ and
consider the initial electron in $n=-1$ state, see the Discussion below.

The wavelength of the laser light is assumed to be much larger than the spacial size $L$
of the $n=-1$ state, so we can neglect spacial variation of the electric field
in the laser pulse. We take the perturbing potential due laser light in the form
\begin{equation} \label{H_W}
 \hW(t) = -ey{\cal E}_0e^{-(2\ln2) t^2/\tau^2}\cos(\omega_Lt),
\end{equation}
where $e$ is the electron charge, $\tau$ is the pulse duration (FWHM),  $\omega_L=2\pi c/\lambda_L$ is
the laser frequency (being of the order of $3\times$10$^{15}$s$^{-1}$),
and ${\cal E}_0$ is the amplitude of electric field.
A Gaussian shape of the laser pulse is widely used in optical experiments and it parameterizes
effectively a profile of electric field in the laser beam.

As a result of a laser shot, the initial state of the system $\Phi_k(t)=\psi_ke^{-iE_kt/\hbar}$
evolves into the final state $\Psi_k(t)=\sum_j c_j(t)\psi_ke^{-iE_jt/\hbar}$, which is
a combination of the eigenstates of $\hH_1$ with suitably chosen coefficients $c_j(t)$.
The resulting time-dependent dipole moment is $\bm D(t)=e\langle \Psi_k(t)|\bm r|\Psi_k(t)\rangle$.

The total Hamiltonian, including the perturbation due to the laser light, is
\begin{equation} \label{H_H}
 \hH = \hH_1 + \hW(t).
\end{equation}
The corresponding time-dependent wave functions are $\Psi_k(t)=e^{-i\hH t/\hbar}\Psi_k(0)$, and
the dipole moment is
\begin{eqnarray} \label{H_D}
\bm D(t)&=& e\langle \Psi_k(0) e^{i\hH t/\hbar}|\hat{\bm r}|e^{-i\hH t/\hbar} \Psi_k(0)\rangle \nonumber \\
        &=& e\langle \Psi_k(0)|\hat{\bm r}(t)|\Psi_k(0)\rangle = e\langle \bm r(t) \rangle.
\end{eqnarray}
Here $\bm r(t)$ is the electron position in the Heisenberg picture.
Thus the dipole moment $\bm D(t)$ is proportional to the time-dependent position
averaged over the electron wave packet.

A time-dependent dipole moment is a source of electromagnetic radiation.
We treat the radiation classically \cite{BohmBook} and take
the radiated transverse electric field to be \cite{JacksonBook}
\begin{equation} \label{H_E}
 {\cal \bm E}_{\perp}(\bm r, t) = \frac{\ddot{\bm D}(t)}{4\pi\epsilon_0c^2}
                           \frac{\sin(\theta)}{R},
\end{equation}
where $\epsilon_0$ is the vacuum permittivity, $\theta$ is an angle between the direction of
electron motion and a position of the observer $\bm R$.
Since $\ddot{\bm D}(t)=e\langle \ddot{\bm r}(t)\rangle$,
Eq. (\ref{H_E}) relates the electric field of the dipole with the average acceleration of the packet.
If the electric field is measured directly by an antenna, one measures the trembling
motion of the wave packet. If the square of electric field is measured in emission or absorption experiments,
the signature of ZB is the existence of peaks corresponding to interband and intraband frequencies and
their dependence on packet's parameters. Accordingly, in the time resolved luminescence experiments
it should be possible to detect directly the motion of the packet with interband and intraband frequencies.

\section{Emitted Electric field}
Now we calculate explicitly the electric field emitted by the trembling electron.
If we consider the perturbation of Eq. (\ref{H_H}), the standard time-dependent perturbation
theory gives for the wave function \cite{LandauBook}
\begin{equation} \label{E_def_Psit}
\Psi_n(t) \simeq \psi_n e^{-iE_nt/\hbar} + \sum_j c_{nj}^1(t)\psi_je^{-iE_jt/\hbar} + \ldots,
\end{equation}
with
\begin{equation} \label{E_def_ck}
 c_{nj}^1(t) = \frac{1}{i\hbar} \int_{-\infty}^t \hW_{nj}(t')e^{i(E_n-E_j)t'/\hbar}dt'.
\end{equation}
Setting the initial state to be $n=-1$ and using Eq. (\ref{H_W}) we have
\begin{equation} \label{E_ck}
 c_{-1j}^1(t) = -\frac{1}{i} \left(\frac{\sqrt{\pi}eL{\cal E}_0\tau}{\hbar\sqrt{\alpha}} \right)
    a_{-1j}^y\ b_{-1j}(t),
\end{equation}
where
\begin{equation}  \label{E_defa} a_{-1j}^y= \frac{1}{L}\int \psi_{-1}^{\dag}(\bm r) y \psi_{j}(\bm r) d \bm r, \end{equation}
\begin{equation} \label{E_defb}
b_{-1j}(t)= \frac{\sqrt{\alpha}}{\sqrt{\pi}\tau}\int_{-\infty}^{t} e^{-\alpha t'^2/\tau^2} e^{i\omega_{-1j}t'}
\cos(\omega_Lt')dt'.
\end{equation}
Here $\omega_{-1j}=(E_{-1}-E_j)/\hbar$ and $\alpha=2\ln 2\approx 1.34$.
Since the size of $n=-1$ state is of the order of $L$, the coefficients $a_{-1j}^y$ are of the order
of unity. It is also is easy to show that $|b_{-1j}| \leq 1$. The dimensionless perturbation expansion
parameter is  $\zeta = \sqrt{\pi}eL{\cal E}_0\tau/\hbar\sqrt{\alpha}$.

To obtain the coefficients $a_{-1j}$ from Eq. (\ref{E_defa})
we calculate the matrix elements of $y$ between
different eigenstates $\psi_n(\bm r)$ of $\hH_1$. The selection rules for
$\langle \psi_n|y|\psi_j\rangle$ of Eq. (\ref{E_defa})
are $|n|-|j|=\pm 1$, so for $n=-1$ there are three non-vanishing
matrix elements corresponding to $j=0,\pm 2$, see Fig. {\ref{Fig0}}.
The approximate wave function $\Psi_{-1}(t)$ is then
\begin{equation} \label{E_Psit}
\Psi_{-1}(t)\simeq e^{-iE_{-1}t/\hbar}\psi_{-1} +
       \sum_{j=0,\pm 2} c_{-1j}^1\ e^{-iE_{j}t/\hbar}\psi_{j}.
\end{equation}
Therefore, the laser shot creates a non-stationary wave packet given by Eq. (\ref{E_Psit}).
This wave packet contains  states with positive and negative energies, which is a
necessary condition to create the ZB motion \cite{Schliemann05,Rusin07b}.
If the packet contains only positive or only negative energy states, the ZB will not occur.
To calculate $\bm D(t)$ we average $\hat{y}$ and $\hat{x}$ over the wave function (\ref{E_Psit}).
As a result we obtain 16 terms, of which one term does not depend on $c_{-1j}$, six terms are
proportional to $c_{-1j}$, and the remaining nine terms are of the second order in $c_{-1j}$.
Since the zero order term does not depend on time, we concentrate on the time-dependent terms
of the lowest order in ${\cal E}_0$, and we have
\begin{equation} \label{E_Dt0}
D_y(t) \simeq {\rm const}+ e L \!\!\! \sum_{j=0,\pm 2}\!\!\!
      c_{-1j}^1\ a_{-1j}^y\ e^{i\omega_{-1j}t} + {\rm h. c.}  + \ldots,
\end{equation}
and similarly for $D_x(t)$, with $a_{-1j}^y$ replaced by $a_{-1j}^x$.
Because the pulse duration $\tau$ is much shorter than the period of ZB oscillations
$T_Z\approx 2\pi/\Omega$, we approximate $b_{-1j}(t)$ by $b_{-1j}(\infty)$. Then [see Eq. (\ref{E_defb})]
\begin{equation} \label{E_binf}
b_{-1j} \approx b_{-1j}(\infty) = \frac{1}{2} \sum_{s=\pm 1} e^{-(\omega_{-1j} +s\omega_L)^2\tau^2/4\alpha}.
\end{equation}
Within this approximation we have
\begin{eqnarray} \label{E_Dt}
D_y(t)&= d_0& \left(-\ \frac{b_{-10}}{2}\sin(\omega^c_0 t)
                    + B^- \ b_{-12}\sin(\omega^Z_1t)
                    \right. \nonumber \\  &&  \left.
                    + B^+\ b_{-1-2} \sin(\omega^c_1t)\right), \nonumber \\
D_x(t)&= d_0&\left( \frac{b_{-10}}{2}\cos(\omega^c_0t)
                    +\ B^-\ b_{-12} \cos(\omega^Z_1t)
                    \right. \nonumber \\   &&  \left.
                   -\ B^+\ b_{-1-2}\cos(\omega^c_1t)\right),
\end{eqnarray}
where $\omega^c_n=\Omega(\sqrt{n+1}-\sqrt{n})$, $\omega^Z_n=\Omega(\sqrt{n+1}+\sqrt{n})$,
$B^{\pm} = \sqrt{2}/2 \pm 3/4$,
and $d_0=-eL\zeta$. Taking the second time derivative of the dipole moment we find
the electric field components of the emitted electromagnetic wave
\begin{eqnarray} \label{E_E(t)}
{\cal E}_y(\bm r,t) &= \Xi(\bm r) &\left( \frac{b_{-1 0}}{2}\sin(\omega^c_0t)
                                        + \frac{b_{-1 2}}{4}\sin(\omega^Z_1t) \right. \nonumber \\
               && \left.                -\ \frac{b_{-1-2}}{4}\sin(\omega^c_1t)\right), \nonumber \\
{\cal E}_x(\bm r,t) &= \Xi(\bm r) &\left(-\ \frac{b_{-1 0}}{2}\cos(\omega^c_0t)
                                         +\frac{b_{-1 2}}{4}\cos(\omega^Z_1t) \right. \nonumber \\
               &&                  \left.+\frac{b_{-1-2}}{4}\cos(\omega^c_1t)\right),
\end{eqnarray}
where $\Xi(\bm r) = d_0\Omega^2 \sin(\theta)/(4\pi\epsilon_0 c^2 R^2)$.
Equations (\ref{E_Dt}) and (\ref{E_E(t)}) are among the main results of our work.
They state that both the induced dipole moment and the corresponding
electric field oscillate with three frequencies.
The frequency $\omega_{-12}=(\sqrt{2}\!+1\!)\ \Omega$ corresponds to the Zitterbewegung,
i.e., to the motion of the packet with an interband frequency.
This frequency corresponds to the interband ZB frequency $\omega_Z=2m_ec^2/\hbar$ of
relativistic electrons in a vacuum. The interband frequency is characteristic of ZB
because the trembling motion occurs due to an interference of electron states
related to positive and negative electron energies \cite{BjorkenBook,GreinerBook}.
The second frequency $\omega_{-1-2}=(\sqrt{2}-1)\ \Omega$ describes the intraband cyclotron motion
of the packet. The third frequency $\omega_{-10}=\ \Omega$ has both interband and intraband
character (see Fig. \ref{Fig0}).
Contrary to the relativistic quantum mechanics, in zero-gap materials like graphene
the interband ZB frequency is not vastly larger than the cyclotron frequency.

\begin{figure}
\includegraphics[width=8.5cm,height=8.5cm]{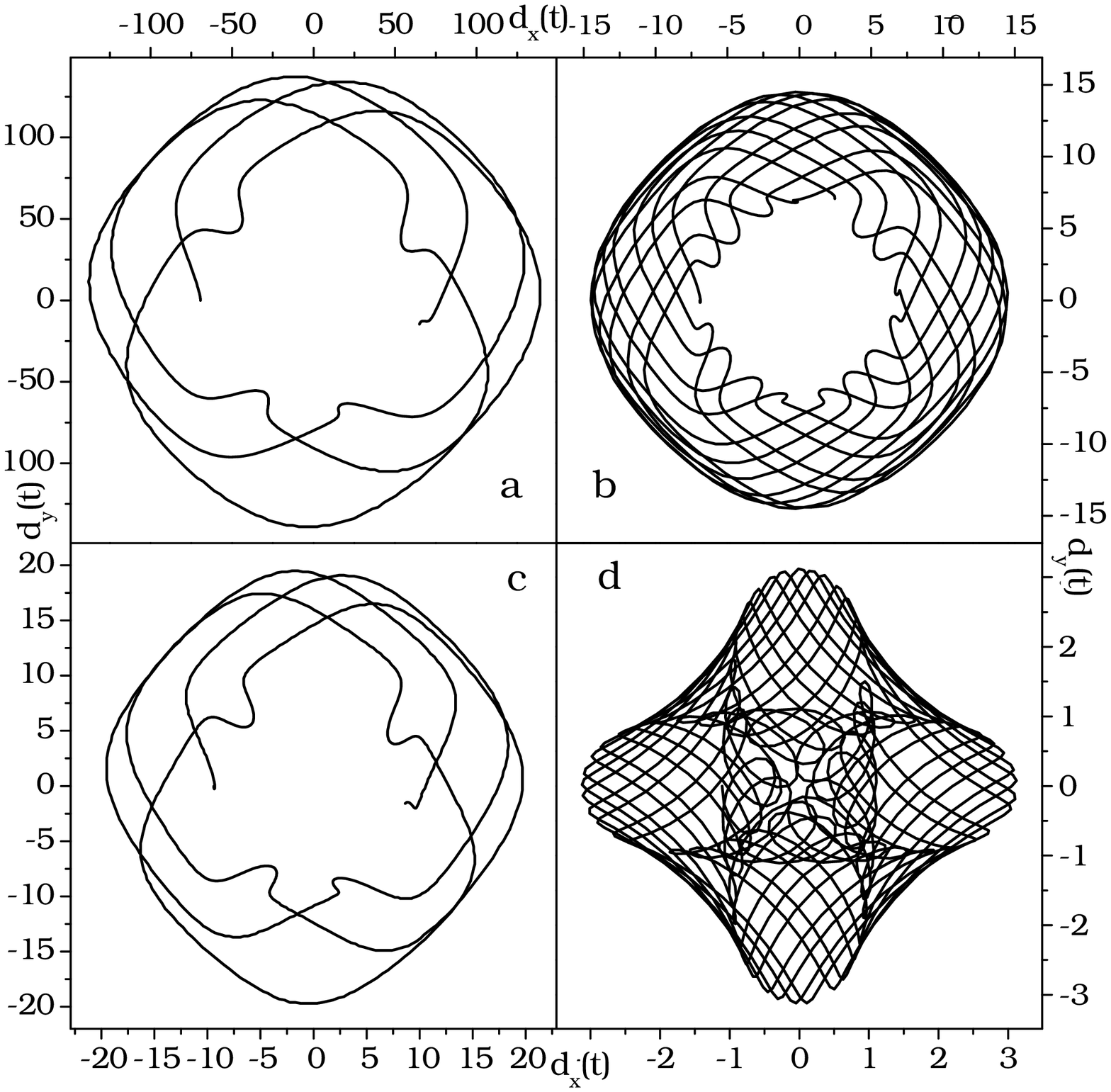}
\caption{Oscillations of dipole moment during the first 1000 fs of electron motion
after the laser pulse. Experimental characteristics: pulse intensity $1\times$ 10$^9$ W/cm$^2$,
 a) $\tau=1.6$ fs, $B=1$ T, b) $\tau=1.6$ fs, $B=10$ T,
 c) $\tau=3.0$ fs, $B=1$ T, d) $\tau=3.0$ fs, $B=10$ T.
Dipole moments in a) and b) are in 10$^{-28}$ [Cm] units,
while in c) and d) they are in  10$^{-31}$ [Cm] units.
The above results refer to very narrow Landau levels, disregarding
broadening due to electron scattering and the e-e interaction, see Section IV. }  \label{FigA}
\end{figure}

\begin{figure}
\includegraphics[width=8.5cm,height=11.33cm]{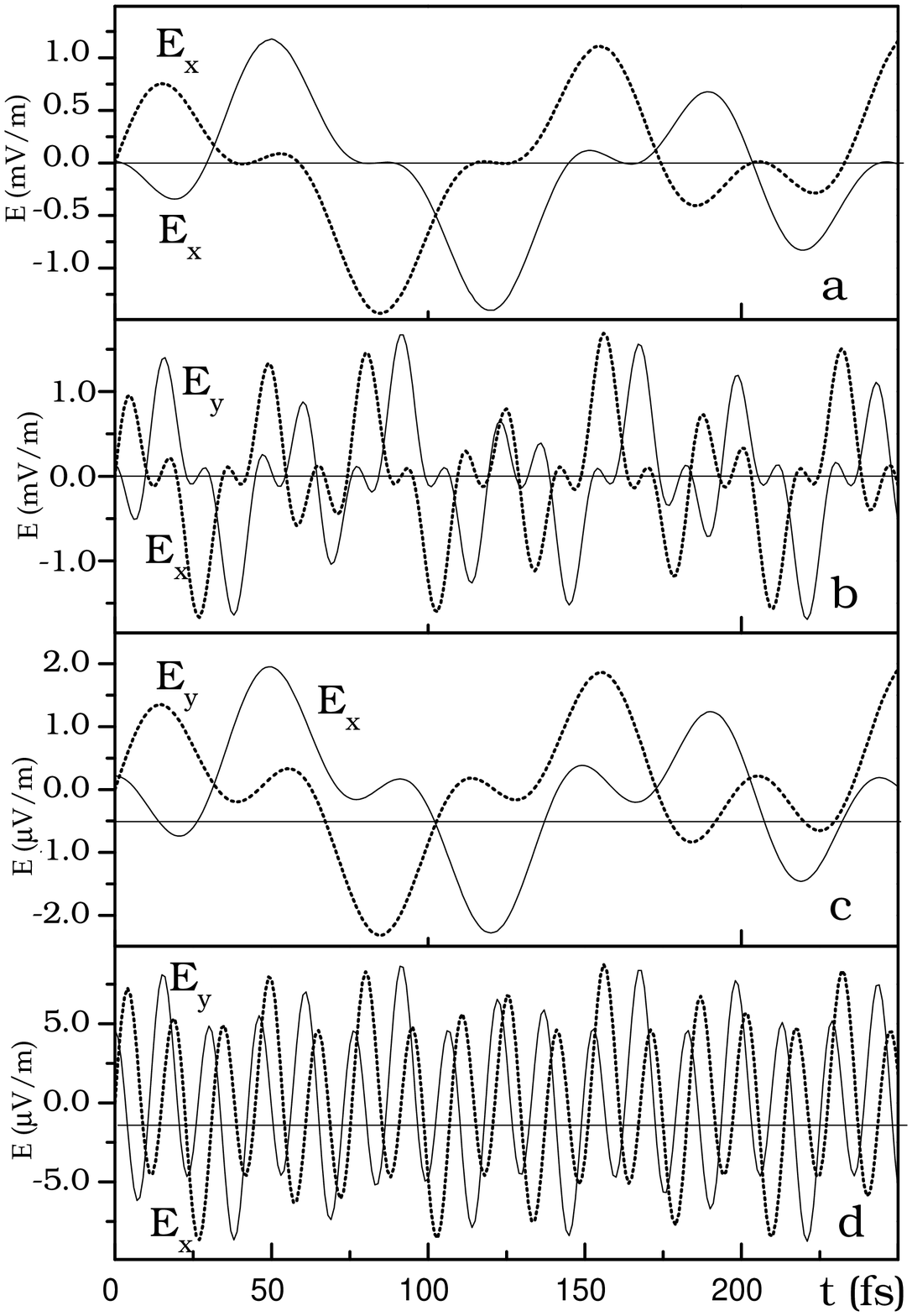}
\caption{Electric field emitted by one electron during the first 250 fs of electron motion
after the laser pulse.  Experimental characteristics: pulse intensity $1\times$ 10$^9$ W/cm$^2$,
 a) $\tau=1.6$ fs, $B=1$ T, b) $\tau=1.6$ fs, $B=10$ T,
 c) $\tau=3.0$ fs, $B=1$ T, d) $\tau=3.0$ fs, $B=10$ T. Note the difference in magnitudes
for $\tau=1.6$ fs and $\tau=3.0$ fs.
The above results refer to very narrow Landau levels, disregarding
broadening due to electron scattering and the e-e interaction, see Section IV.  }  \label{FigB}
\end{figure}

In Fig. \ref{FigA} we plot the oscillating dipole moment within the first 1000 fs of
motion after the laser shot for two magnetic fields $B=1$ T and $B=10$ T, and two laser pulses.
The first pulse (Figs. \ref{FigA}a, \ref{FigA}b) has a duration $\tau=1.6$ fs and a base laser wavelength
$\lambda_L=650$ nm. This pulse has a sub-monocycle duration and it is the shortest
pulse created experimentally within the visible laser wavelength \cite{Shverdin05}.
In Figs. \ref{FigA}c, \ref{FigA}d we assume pulse duration $\tau=3.0$ fs
and a laser wavelength $\lambda_L=720$ nm.
This pulse has 1.25 of the laser monocycle and its experimental
properties were discussed in Ref. \cite{Cavalieri07}.
The use of a few monocycle pulses (with $\tau > 5$ fs) is not effective,
since the probability of excitation of a wave packet in Eq. (\ref{E_Psit}) drops
exponentially with pulse duration $\tau$, see Eq. (\ref{E_binf}).

In Fig. \ref{FigB}  we plot the corresponding electric field for the same parameters
during the first 250 fs of oscillations.
We assume the laser intensity to be $I=1.0 \times$ 10$^9$ W/cm$^2$, the emitted
electromagnetic wave detected at the angle $\theta=45^{\rm o}$, and the distance $R=1$ cm.
All the quantities in Figs. \ref{FigA} and \ref{FigB} are calculated per one electron.
Since the frequencies are incommensurable, the electron trajectories $\bm r(t)$
are not closed and there is no repeated pattern of oscillations.
The motion of the wave packet is permanent in the time scale
of femtoseconds or picoseconds but there is damping of the motion due to
the light emission in a long time scale.
The results shown in Figs. \ref{FigA} and \ref{FigB} refer to very narrow Landau levels,
disregarding broadening due to electron scattering and the e-e interaction, see Section IV.

We can draw the following qualitative conclusions from Figs. \ref{FigA} and \ref{FigB}.
First, for small magnetic fields $B$ the period of oscillations
is longer than for large fields,  which is related to the basic frequency $\Omega$.
Second, irrespective of the variation of $\Omega$ with $B$,  for small fields the oscillations
are dominated by the low  (cyclotron) frequency, while at stronger $B$ the high (ZB) frequency dominates.
Finally, comparing the magnitudes of dipole moment or emitted electric field for $\tau=1.6$ fs
with the corresponding values for $\tau=3.0$ fs we observe that  the amplitude of
oscillations depends very strongly on the duration $\tau$ of the pulse.

\begin{table}
\begin{tabular}{|c|c|c|c|c|}
 \hline
 $\tau$ (fs)&B (T)&$b_{-1-2}$             & $b_{-10}$             & $b_{-12 }            $\\ \hline
            &  1  & $2.07\times$10$^{-2}$ & $2.09\times$10$^{-2}$ & $2.18\times$10$^{-2}$ \\
  1.6       & 10  & $2.10\times$10$^{-2}$ & $2.26\times$10$^{-2}$ & $3.20\times$10$^{-2}$ \\
            & 40  & $2.20\times$10$^{-2}$ & $2.84\times$10$^{-2}$ & $6.96\times$10$^{-2}$ \\ \hline
            &  1  & $1.52\times$10$^{-5}$ & $1.65\times$10$^{-5}$ & $2.45\times$10$^{-5}$ \\
  3.0       & 10  & $1.76\times$10$^{-5}$ & $3.22\times$10$^{-5}$ & $1.89\times$10$^{-4}$ \\
            & 40  & $2.63\times$10$^{-5}$ & $1.13\times$10$^{-4}$ & $2.74\times$10$^{-3}$ \\ \hline
\end{tabular}
\caption{Coefficients $b_{-1j}$ for the electric field in Eq. (\ref{E_E(t)}) for different pulse
         durations $\tau$ and various magnetic fields $B$. }
\end{table}

To analyze these effects quantitatively we collected in Table 1 the coefficients $b_{-1j}$
of Eq. (\ref{E_binf}), used for the calculation of electric field in Eq. (\ref{E_E(t)}).
The results presented in Table 1 show that for fixed $B$ and  $\tau=1.6$ fs
the coefficients $b_{-1-2}$ and $b_{-10}$
are nearly three orders of magnitude larger than those for $\tau=3.0$ fs.
We also note that, for $B=1$ T, all $b_{-1j}$ are nearly identical,
while for $B=40$ T there are visible differences between various $b_{-1j}$.
This difference is a factor of 3 for $\tau=1.6$ fs, while for $\tau=3.0$ fs
the coefficient $b_{-21}$ is  two orders of magnitude larger than $b_{-1-2}$.
This explains the dominance of the interband ZB frequency
for large $B$ in Figs. \ref{FigA} and \ref{FigB}.
The conclusion from this analysis is that the optimum conditions for observing the ZB, i.e.
the packet motion with both interband and intraband frequencies, is the regime of magnetic
fields of a few Tesla, since in this regime the two kinds of motion exist with comparable weights.
One should note that the coefficients $b_{-1j}$, as defined in Eq. (\ref{E_defb}),
are closely related to the coefficients $c_{-1j}$  in the perturbation series,
see Eqs. (\ref{E_ck}) and (\ref{E_binf}). A practical lower limit for magnetic fields
is the condition that an energy distance between LLs should be larger than
the widths $\Gamma_L$ of LLs in graphene. According to Ref. \cite{Sadowski06}
one observes resonant magneto-optical transitions, both interband and intraband, in
graphene beginning with a magnetic field of $B \approx 0.4$ T. Thus
$B \approx 0.5$ T seems to be the lowest possible magnetic field suitable
for the experiment described above. The Landau level broadening and its effect on ZB
are discussed in the next section.

Comparing Figs. \ref{FigA} and \ref{FigB}  for fixed $\tau$ one observes a strong dependence
of the dipole moment on the magnetic field, while the radiated electric field
depends only weakly on $B$. The reason is that the electric
field of Eq. (\ref{E_E(t)}) is multiplied by a factor $d_0\Omega^2$.
Because $d_0=-eL\zeta$, $\zeta \propto L$, and $\Omega = \sqrt{2}u/L$,
the  product $d_0\Omega^2$ does not depend on $B$. On the other
hand, $d_0 \propto L^2 \propto B^{-1}$, and the dipole moment $\bm D(t)$ depends strongly on $B$.
Therefore, the electric field of the emitted electromagnetic wave (and the radiated power)
does not change significantly with the magnetic field intensity, it
depends on $B$ only via coefficients  $b_{-1j}$, see Table 1.

For $B=1$ T, the basic frequency $\Omega$ is $5.41\times$ 10$^{13}$ s$^{-1}$,
which corresponds to $f_{\Omega}=\Omega/(2\pi)=8.61$ THz.
The cyclotron frequency is $f_c=3.53$ THz, while the ZB frequency is
$f_Z=20.8$ THz. All the three frequencies are within the range of currently
available THz photoconductive antennas, see e.g. Ref. \cite{Kono01}, and it should be
possible to detect the emitted field experimentally.
In contrast, for $B=40$ T the corresponding frequencies are $f_{\Omega}=54.5$ THz, $f_c=22.3$ THz and
$f_Z=131$ THz, and they are more difficult to detect.
This is the other reason for using low magnetic fields in the experiment.

\section{ZB in real samples}
In the previous section we considered an idealized case of very narrow Landau levels in graphene.
In real samples additional effects occur and their presence affects
the motion of the wave packet.
Two effects may play a role in the proposed experiment:
the electron - electron (e-e) interaction \cite{Li09} and  the presence of disorder \cite{Zhu09,Stampfer09}.
The scanning tunnelling spectroscopy results of Ref. \cite{Li09} indicate that
the e-e interaction leads to a Lorentzian shape of DOS of the Landau levels and it
opens an energy gap between the electrons and holes. Thus the massless Dirac fermions acquire a small
non-zero mass. As shown in the numerical simulations of Ref. \cite{Zhu09}, the presence of disorder
changes the shape of DOS from Lorentzian to Gaussian peaks.
Additionally, the disorder potential may change the position of the Fermi level within sample.

The band-gap caused by the e-e interaction is of the order of $10$ meV \cite{Li09},
so at a magnetic field of $B=1$ T it is much smaller than the basic energy
$\hbar \Omega \approx 36$ meV. In this case the energy spectrum of graphene in a magnetic
field is described by an analogue of the Dirac equation, whose energy levels and eigen-functions
are well known. The trembling motion of the packet  will not change qualitatively, as compared to the
above description, but it will oscillate with the interband frequency
$\tilde{\Omega}=\sqrt{\Omega^2+E_{gap}^2/\hbar^2}$, and all frequencies will have slightly
different values than those calculated in the gapless model.

On the other hand, the broadening of the Landau levels may strongly influence the trembling
motion of the wave packet. To analyze the overall impact of all the effects leading to
the level broadening: disorder, e-e interaction, electron-phonon scattering,  etc.,
we assume finite widths of all energy levels, characterized by broadening parameters $\Gamma_n$.
We treat $\Gamma_n$ as phenomenological quantities determined experimentally and
including all scattering mechanisms existing in real samples.
We approximate the broadening of DOS by a Lorentzian line-shape \cite{Li09} irrespective of
the detailed scattering mechanism \cite{Zhu09}. In this approximation the Landau energies $E_n$
are replaced by complex energies $\tilde{E_n} = E_n + i\Gamma_n$. After the replacement
the dipole moment of Eq. (\ref{E_Dt0}) changes to
\begin{equation} \label{E_DtG0}
D_y^{\Gamma}(t) \simeq {\rm const}+ e L \!\!\! \sum_{j=0,\pm 2}\!\!\!
      c_{-1j}^1\ a_{-1j}^y\ e^{i\omega_{-1j}t-\Gamma_jt} + {\rm h. c.}  + \ldots,
\end{equation}
which leads to [cf. Eq. (\ref{E_Dt})]
\begin{eqnarray} \label{E_DtG}
D_y^{\Gamma}(t)&= d_0& \left(-\ \frac{b_{-10}}{2}\sin(\omega^c_0 t)e^{-\Gamma_{0}t}
                    \right. \nonumber \\  &&  \left.
                    + B^- \ b_{-12}\sin(\omega^Z_1t)e^{-\Gamma_{+2}t}
                    \right. \nonumber \\  &&  \left.
                    + B^+\ b_{-1-2} \sin(\omega^c_1t)e^{-\Gamma_{-2}t} \right), \nonumber \\
D_x^{\Gamma}(t)&= d_0&\left( \frac{b_{-10}}{2}\cos(\omega^c_0t)e^{-\Gamma_{0}t}
                    \right. \nonumber \\  &&  \left.
                    +\ B^-\ b_{-12} \cos(\omega^Z_1t)e^{-\Gamma_{+2}t}
                    \right. \nonumber \\   &&  \left.
                   -\ B^+\ b_{-1-2}\cos(\omega^c_1t)e^{-\Gamma_{-2}t} \right).
\end{eqnarray}
As to the numerical values of $\Gamma_n$, we take after Ref. {\cite{Li09}} $\Gamma_{\pm 2}= 5.1$ meV.
For the state $n=0$ we use $\Gamma_{0}= 4.3$ meV \cite{Zhu09} which corresponds to the disorder potential
$V_g=120$ meV \cite{Stampfer09}. These values are similar to the line widths of $\Gamma_n = 7$ meV
measured at $B=1$ T in far infrared transmission experiments \cite{Orlita08}, and
$\Gamma_n=1.6$ meV determined by the Quantum Hall Effect at $B \sim 40$ T \cite{Zhang06}.

\begin{figure}
\includegraphics[width=8.5cm,height=8.5cm]{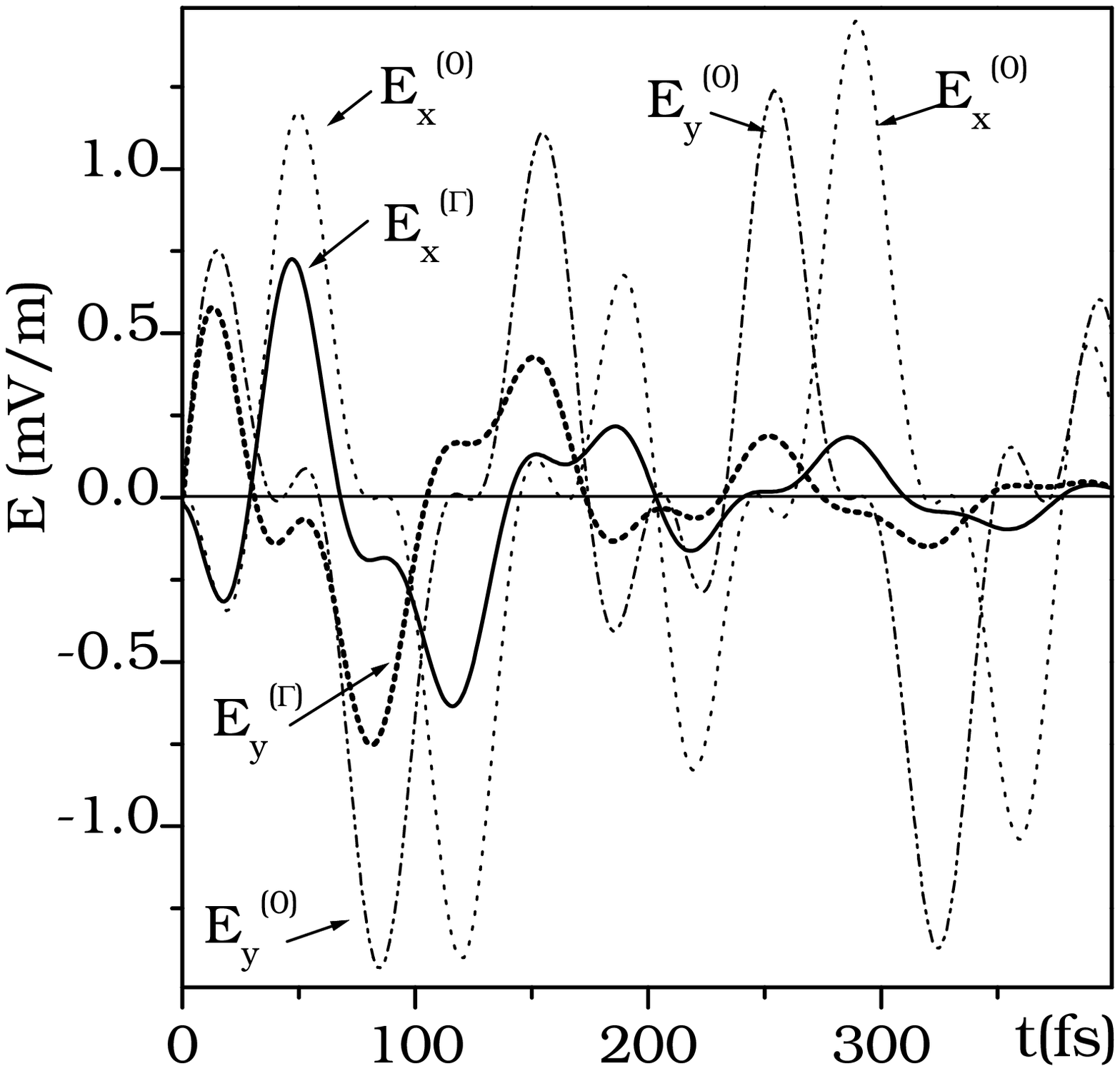}
\caption{Calculated electric fields $E_x(t)$ and $E_y(t)$ emitted by one trembling electron during
the first 400 fs after the laser pulse.
Pulse parameters: intensity $1\times$ 10$^9$ W/cm$^2$, $\tau=1.6$ fs, magnetic field $B=1$ T.
Bold lines - electric fields for broadened Landau levels described by Lorentzian
line-shapes with experimental values of $\Gamma_n$ (see text). Thinner  lines - electric fields for
delta-like Landau levels. The latter results are the same as those shown in Fig. \ref {FigB}a. }
 \label{FigE}
\end{figure}

The electric field $\bm E(t)$ is calculated, as before, as a second time derivative of $\bm D^{\Gamma}(t)$,
see Eq. (\ref{E_Dt}). In Fig. \ref{FigE} we plot the electric field emitted
by an oscillating electron within the first 400 fs of
motion after the laser shot of the width $\tau=1.6$ fs in a magnetic field $B=1$ T.
The two bold lines describe calculated electric fields  $E_x(t)$ and $E_y(t)$
for the Landau levels having the broadening parameters $\Gamma_n$ indicated above.
The two thinner lines show the electric fields calculated without damping ($\Gamma_n=0$),
see Eq. (\ref{E_Dt}) and Fig. \ref{FigB}a.
Within the first 50 fs of motion the electric fields emitted in the two cases are similar,
but later the damping of the emitted fields for broadened levels is visible.
After around 400 fs the trembling motion in real case disappears.
It can be seen that the maxima of oscillations for the damped ZB motion coincide with the undamped ones.
The general conclusion from Fig. \ref{FigE} is that the existence of disorder,
many-body effects or other scattering mechanisms changes the persistent ZB motion to a decaying one,
within the characteristic lifetimes for these processes: $\tau_n=1/\Gamma_n \approx 130$ fs.
Nevertheless, since the parameters $\Gamma_n$ used in the calculations correspond
to the measured lifetimes in real graphene samples, it follows that the broadening of the Landau
levels does not prevent the existence of ZB. Clearly, a lower disorder in better samples will
result in longer decay times for ZB.

\section{Time-resolved Luminescence}
\begin{figure}
\includegraphics[width=8.5cm,height=8.5cm]{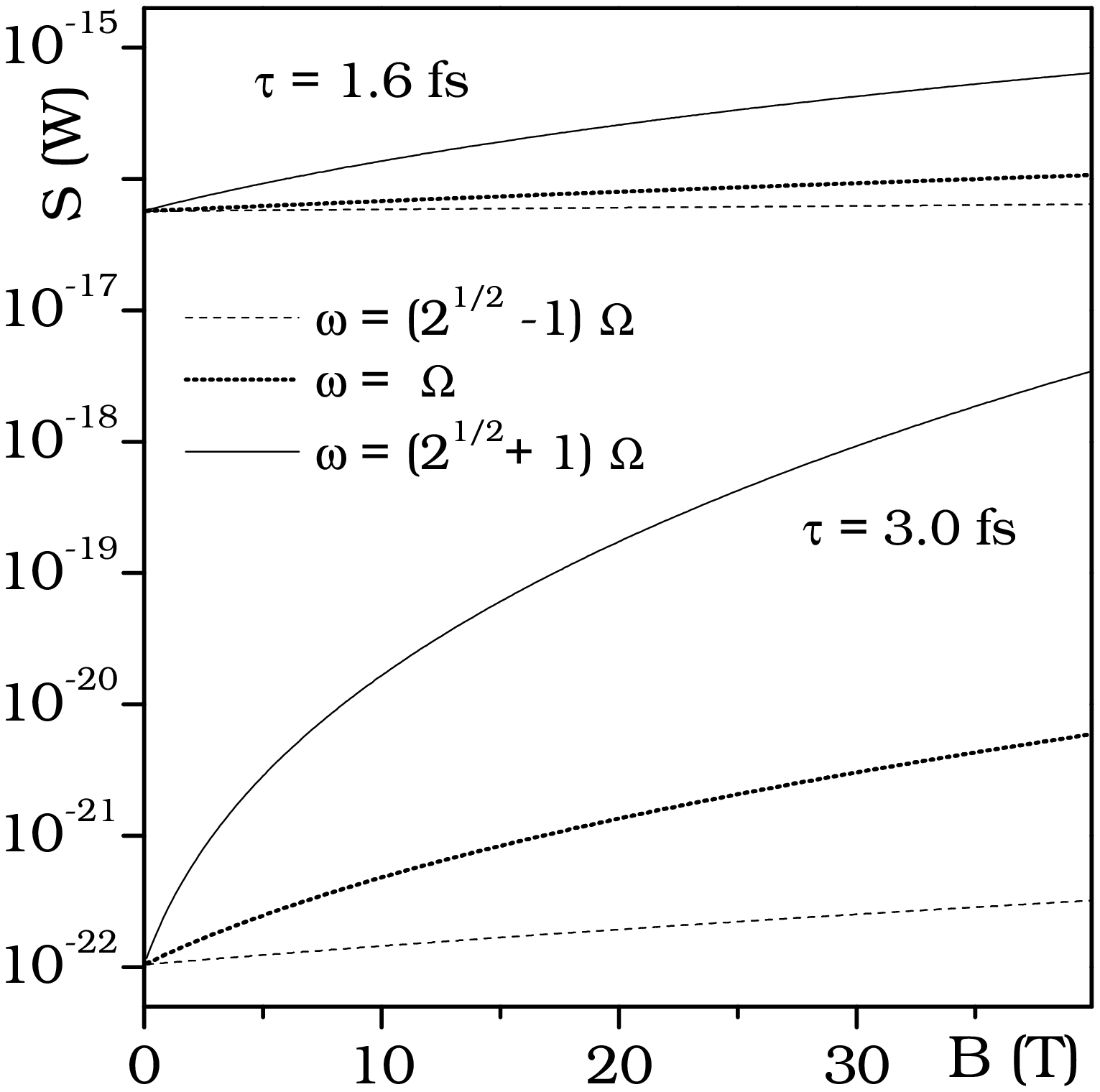}
\caption{Intensity of three emission lines versus magnetic field for two laser pulses
        of different durations $\tau$.} \label{FigC}
\end{figure}

Knowing the electric field emitted by an oscillating dipole we can calculate the intensity of
radiated light. If ${\cal E}(t)$ is given by a sum of cosine (or sine) functions:
${\cal E}(t) = \sum_j f_j\cos(\omega_jt)$, the emitted power averaged over
a long time is $\bar{P} = (1/2)\sum_j|f_j|^2$. The total power $S$ passing through
a closed spherical surface at a distance $R$ from the sample can be calculated
integrating $\bar{P}$ over the enclosing surface.
Using Eq. (\ref{E_E(t)}) and the above formula we find
\begin{equation} \label{S_S}
 S = \frac{d_0^2\Omega^4}{96\pi\epsilon_0c^3} \left(4|b_{-10}|^2 + |b_{-12}|^2 + |b_{-1-2}|^2 \right),
\end{equation}
which is the Larmor formula for our problem.
The emitted spectrum $S$ consists of three lines of different intensities having the
frequencies $\omega = \Omega$, $\omega=(\sqrt{2}+1)\Omega$ and $\omega=(\sqrt{2}-1)\Omega$.
The existence of lines with interband frequencies is the signature of Zitterbewegung in this system.
In Fig. \ref{FigC} we plot the relative light intensities for different emission lines versus magnetic field $B$
on the logarithmic scale. The upper curves correspond to  $\tau=1.6$ fs, the lower ones to $\tau=3.0$ fs.
In both cases the intensities depend on magnetic field, this dependence is most pronounced
for the ZB frequencies. However, for $\tau=1.6$ fs the dependence of intensities on
the magnetic field is much weaker than for $\tau=3.0$ fs. At high $B$  the spectrum is dominated
by the ZB frequencies. It should be noted  that the intensity of radiation is proportional to $\Omega^4$ and to
$\zeta^2\propto {\cal E}_0^2$, as for the Thompson scattering.

\begin{figure}
\includegraphics[width=8.5cm,height=8.5cm]{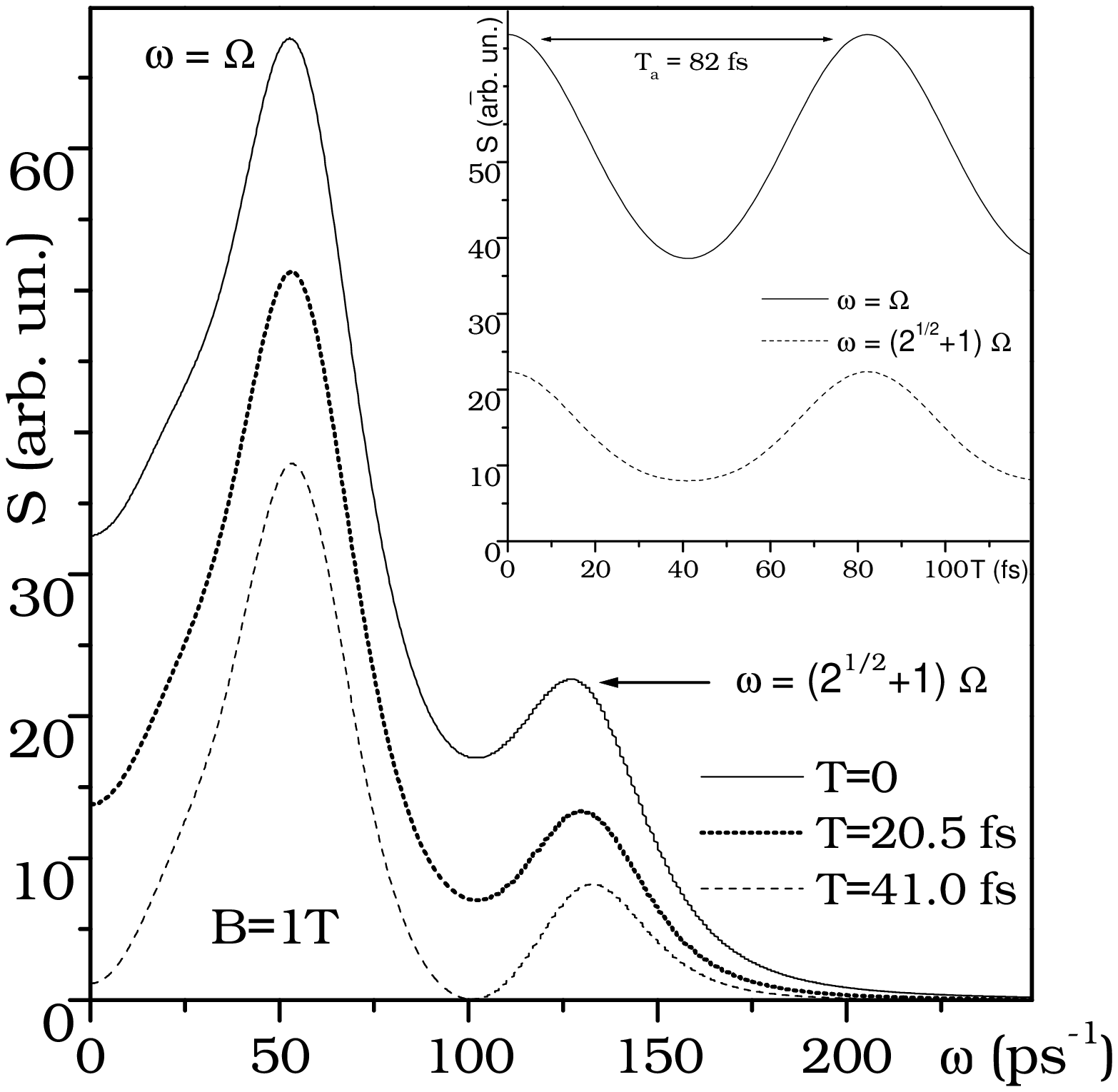}
\caption{Calculated power spectra of the emitted light for different opening gate times $T$.
        For the assumed resolutions $\gamma=3$ cm$^{-1}$ and $\Gamma=1000$ cm$^{-1}$
        the peaks at $\omega = \Omega$  and $\omega = (\sqrt{2}-1)\ \Omega$ are not resolved.
Inset: Intensity of two peaks:  $\omega=\Omega$ and $\omega=(\sqrt{2}+1)\Omega$ versus gate time $T$.
       Repetition of oscillation pattern occurs after $T_a = \sqrt{2}\ \Omega = 82.2$ fs for $B=1$ T.} \label{FigD}
\end{figure}

To observe the motion of electron represented by a wave packet
one can use the time-resolved luminescence \cite{Eberly77,Kowalczyk90,Dunn93}.
In this technique, the electric field $\cal{\bm E}$(t) of the light emitted by a sample
is transmitted through two filters:
a time gate $B(t,T)$ \begin{equation} B(t,T) = \exp(-\Gamma|t-T|), \end{equation}
and a frequency filter $H(t,\omega_F)$
\begin{equation} H(\omega,\omega_F) = \frac{\gamma^2}{\gamma^2 +(\omega-\omega_F)^2}. \end{equation}
The time gate lets the field pass for $ T-\Gamma/2 < t < T+\Gamma/2 $, where $\Gamma$ is the gate's parameter
describing the width of the window, and $T$ is the center of the window.
The frequency filter selects frequencies close to $\omega_F$ with the resolution $\gamma$.
The transmitted field is
\begin{equation} \label{S_EHB}
  {\cal \bm E}_{HB}(t) = \int_{-\infty}^t H(t-t',\omega_F)B(t',T){\cal \bm E}(t')  dt',
\end{equation}
where $H(t,\omega_F)$ is the Fourier transform of $H(\omega,\omega_F)$.
The time-dependent spectrum of ${\cal \bm E}(t)$ is
\begin{equation}  \label{S_defS}
  S(T,\omega_F) = \int dA  \int_{-\infty}^{\infty}  \left|{\cal \bm E}_{HB}(t)\right|^2 dt,
\end{equation}
where the first integration is over the sphere of radius $R$ enclosing the sample.
For ${\cal \bm E}(t)$ given by Eq. (\ref{E_E(t)}) we obtain
\begin{equation} \label{S_S(t)}
  S(T,\omega_F)= M \sum_{\alpha, \beta, r}  b_{\alpha}^r  b_{\beta}^r
     e^{-i(\omega_{\alpha}-\omega_{\beta})T} G(\alpha,\beta),
\end{equation}
where M is a constant and $\alpha, \beta = (n,j,s)$. The summation is restricted to
$n=-1$, $j=0,\pm 2$, and $s=\pm 1$. The index $r=x,y$, and we
denote $\omega_{\alpha},\omega_{\beta} \equiv s\omega_{nj}$.
Factors $b_{\alpha}^r$  and $b_{\beta}^r$ are the coefficients in front of trigonometric
functions in Eq. (\ref{E_E(t)}). We have
$b^y_{-10}=b_{-10}/2$, $b^y_{-12}=b_{-12}/4$, and $b^y_{-1-2}=-b_{-1-2}/4$. Similarly
$b^x_{-10}=-b_{-10}/2$, $b^x_{-12}=b_{-12}/4$, and $b^x_{-1-2}=b_{-1-2}/4$.
The analytical function $G(\alpha,\beta)$ is defined by Eq. (17) of Ref. \cite{Kowalczyk90}.

In Fig. \ref{FigD} we show the calculated time-dependent spectrum of ZB oscillations
for the following parameters:
laser pulse width $\tau=1.6$ fs, laser wavelength $\lambda_L=650$ nm, magnetic field $B=1$ T,
frequency  filter $\gamma=3$ cm$^{-1}$, and the gate width $\Gamma=1000$ cm$^{-1}$. Three curves correspond
to frequency spectrum observed at three gate opening times $T$.
If follows from Eq. (\ref{S_S(t)}) and from the selection rules for $j$, see Eq. (\ref{E_defa}),
that the time dependent spectrum $S(T,\omega_F)$ has maxima at the frequencies: $\omega=\ \omega_c$,
$\omega=\ \Omega$, and $\omega=\ \omega_Z$. Because of finite resolutions of the time gate
and the frequency filter, the maxima for $\omega=\ \omega_c$ and $\omega=\ \Omega$ form one unresolved peak.

Changing the gate opening time $T$ one should observe
different frequency spectra.  The spectrum repeats its pattern after $T_a=2\pi/\omega_a = 82.2$ fs,
where $\omega_a=\omega_{-10}-\omega_{-1-2}=\sqrt{2}\Omega$.  To illustrate this time dependence
we plot in the inset of Fig. \ref{FigD} intensities of the emitted radiation for two frequencies:
$\omega=\ \Omega$ and $\omega=\ \Omega_Z$ versus gate's opening time $T$. We see that,
for both maxima, the intensities oscillate with the period $T_a$.
The oscillating pattern of $S(T,\omega_F)$ is also the signature of Zitterbewegung.

\section{Discussion}

In section II we assumed that the Fermi level in monolayer graphene coincides with the
Landau level $n=-2$ and considered the initial electron in the state $n=-1$, see Fig. \ref{Fig0}.
This means that, before the electron reacts to a short laser pulse, it must be pumped to
the state $n=-1$ from lower Landau levels. A conventional light source is suitable
for such pumping but, in order to achieve high intensities of the emitted lines, one should
use a laser pump in resonance with $\hbar \omega = E_{-1} - E_{-2}$ energy.
It should be emphasized that the upper component of the state $n=-1$ in a magnetic field
is described by the Gaussian wave function in space, but, since it is an eigenstate,
it does not have a time dependence. The decisive factor is the subsequent laser pulse
which excites a series of electron eigenstates.

The results described by Eq. (\ref{E_E(t)}) are obtained for {\it one}
electron. The population of the Landau level is $eB/h$ and the total intensity of
radiation is obtained by adding the contributions from all
electrons in the initial $n=-1$ level.
In our proposition we select the initial electron state $n=-1$ for several reasons.
Due to the selection rules $\Delta |j|=\pm 1$ there exist three frequencies
$\omega_{-1j}$ contributing to the electron motion.
If the state $n=0$ were selected, there would be only two non-vanishing matrix elements
$y_{0,-1}=y_{0,+1}$ corresponding to the same frequency $\Omega$, see Fig. \ref{Fig0}.
Assuming the Fermi energy at the $n=-2$ LL and supposing, in consequence, the state $n=-2$
to be roughly half filled, we avoid the Pauli exclusion principle in the calculation
of the dipole matrix elements. Therefore, one-particle formalism of ZB calculation can be
applied, see \cite{Krekora2004,Wang08}. The presence of a magnetic field is essential for a
successful experiment since, as we said above, in the absence  of magnetic field
the ZB oscillations have a transient character with the decay time
of tenths of femtoseconds \cite{Rusin07b,Zawadzki08,Demikhovskii08}
and the detection of such oscillations is difficult. However, as we showed in Section IV,
the broadening of Landau levels also leads to a transient character of ZB.
In high quality graphene samples with small disorder the electron ZB should
last longer after the laser pulse.

From the early eighties, short laser pulses were used in quantum chemistry to excite wave
packets in molecules \cite{Zewail99}. After a laser shot, the non-stationary wave packet
evolves in time and its motion is measured in many ways: absorption, luminescence,
Raman scattering, etc., using the so called pump and probe method.
The experiment proposed here is in principle similar to these techniques.
Our additional requirement is the necessity to use monocycle or sub-monocycle
laser pulses in order to excite the electron packet with both positive and
negative energies \cite{GreinerBook}.

In the nineties, the Bloch oscillations were observed in superlattices in a static electric field.
In these experiments the electric field creates a set of
discrete levels (Wannier-Stark ladder, WSL). Then, a short laser pulse creates a wave packet
consisting of many discrete states and an oscillating dipole moment appears.
The oscillations of the dipole moment were measured either by detection
of THz radiation \cite{Martini96}, or by measuring change in the positions of WSL
energies in pump and probe experiments \cite{Lyssenko96}. There exist several similarities
between the Bloch oscillator experiment and our proposition.
In both cases the system is quantized by an external field, the laser
pulse creates an electron wave packet consisting of many discrete levels,
and the oscillations of the wave packet lead to the time-dependent dipole moment
that can be observed experimentally. The main difference is, again, that in order
to observe ZB one needs to create an electron packet having both positive and negative
energy states. Finally, similar techniques to the one proposed here were applied
to observe a coherent emission from double-well systems \cite{Luryi91,Roskos92}.

In our treatment we used the first order time-dependent perturbation theory, see Eq. (\ref{E_Psit}).
For very strong electric fields this approach may be insufficient.
However, the fast decrease of $b_{-1j}$ coefficients with increasing pulse width $\tau$
assures in practice the validity of the perturbation expansion.

It was shown in the previous work on ZB  that its existence
is related to a nonzero momentum of the electron \cite{Schliemann05,Rusin07b}.
In our gauge for $\bm A$, the initial electron state has an initial momentum $k_x$.
A laser pulse $\hat{W}(t)$ creates the state with nonzero momentum also in the $y$ direction.
Direct calculations show that at $t=0$ the average momentum is
$\langle \hat{p}_i \rangle$= $\langle \Psi(0)|(\hbar/i)(\partial/\partial x_i)|\Psi(0)\rangle$
which gives $\langle \hat{p}_y \rangle =2\zeta\hbar/L$ and
$\langle \hat{p}_x \rangle= k_x$. These results are obtained in the lowest order in
${\cal E}_0$. The asymmetry between $x$ and $y$ directions
follows from the electric field ${\cal} E$ directed parallel to the $y$ axis.

As we said above, several conditions should be met in order to observe the ZB.
In our understanding, all these conditions can be fulfilled:
p-doped monolayer graphene, ultrashort pulse of the required intensity,
detection of the emitted electric field in the 3 - 21 THz range, or an
observation of the time resolved spectra. It should be possible to satisfy these conditions
in a single experiment.

\section{Summary}
To summarize, we proposed and described a possible method to observe
the trembling motion of electrons in graphene in a
magnetic field. The central point is that we did not assume anything about the shape
of the initial electron wave packet. We calculated the time dependent dipole moment
induced by an ultra-short laser pulse. For electrons located initially in the $n=-1$
state the induced dipole moment oscillates with three frequencies,
of which the frequency $\omega_Z = (\sqrt{2}+1)\ \Omega$ is the signature of Zitterbewegung.
A possibility of performing such an experiment and detecting the $\omega_Z$ frequency are
discussed and it appears that the current experimental techniques are sufficient
for a successful observation of ZB in high quality graphene samples.

\acknowledgements
This work was supported in part by The Polish Ministry of Science and Higher Education through
Laboratory of Physical Foundations of Information Processing.

\appendix
\section{}
It is well known that in the Brillouin zone of monolayer graphene there exist two inequivalent
minimum points $K_1$ and $K_2$ \cite{Gusynin07,Bena07}. Above we calculated contributions
from an electron at the point $K_1$ defined by Hamiltonian (\ref{H_H}).
The question arises: what is the contribution of an electron at the point $K_2$?
To elucidate this issue we consider the Hamiltonian for
electrons at the $K_2$ point of BZ \cite{Gusynin07,Bena07}
\begin{equation} \label{A aap}
 \hH_2 = \hbar\Omega\left(\begin{array}{cc}  0 & \hap\\ \ha& 0 \\  \end{array}\right).
\end{equation}
Comparing Eq. (\ref{A aap}) with Eq. (\ref{H_H}) it is seen that
 $\hH_2= -\hH_1^T$. The eigenenergies of both Hamiltonians are the same, but
the eigenstates of $\hH_2$ are
\begin{equation} \label{A_phi}
\chi_n(\bm r)= \frac{e^{ik_xx}}{\sqrt{4\pi}}
               \left(\begin{array}{c} \phi_{|n|}(\xi) \\ {\rm sgn}(n) \phi_{|n|-1}(\xi)
               \end{array}\right),
\end{equation}
which differs from $\psi(\bm r)$ of Eq. (\ref{H_psi}) by an exchange of the upper and lower
components and by the change of sign of $\phi_{|n|-1}(\xi)$ state.
The matrix elements are $\langle\psi|y|\psi\rangle$ = $\langle\chi|y|\chi\rangle$, so both Hamiltonians give
the same selection rules. The perturbed wave function $\Upsilon(\bm r,t)$ for
the electron at the $K_2$ point is [see Eq. (\ref{E_Psit})]
\begin{equation} \label{A_Chit}
\Upsilon_{-1}(t)\simeq e^{-iE_{-1}t/\hbar}\chi_{-1} +
       \sum_{j=0,\pm 2} c_{-1j}^1\ e^{-iE_{j}t/\hbar}\chi_{j},
\end{equation}
with the same coefficients $c_{-1j}$ as in Eq. (\ref{E_Psit}). In consequence,
the wave packet at $t=0$ for the electron at the $K_1$ point is [see Eq. (\ref{E_Psit})]
\begin{equation} \label{A_Psi0}
\Psi_{-1}(0)\simeq \psi_{-1} + \sum_{j=0,\pm 2} c_{-1j}^1\ \psi_{j},
\end{equation}
while for the $K_2$ point the initial wave packet is
\begin{equation} \label{A_Chi0}
\Upsilon_{-1}(0)\simeq \chi_{-1} + \sum_{j=0,\pm 2} c_{-1j}^1\ \chi_{j}.
\end{equation}
Thus, the same laser shot creates two {\it different} electron wave packets at $K_1$ and $K_2$ points.
This is in contrast to the assumption made in Ref. \cite{Rusin08a}, where the same wave packet
was assumed for both $K_1$ and $K_2$ points. The result of this assumption
was a partial cancellation of one of the electric current components.
This was unphysical since it violated the rotational symmetry of the $x-y$ graphene plane.
In the present approach, the two wave packets evolve according to {\it different} Hamiltonians.
There is $\Psi_{-1}(t) = e^{-i\hH_1t/\hbar}\Psi_{-1}(0)$ and
     $\Upsilon_{-1}(t) = e^{-i\hH_2t/\hbar}\Upsilon_{-1}(0)$.
A direct calculation shows that now the contributions to the electric current and dipole oscillations
arising from electrons excited at the two nonequivalent points $K_1$ and $K_2$ are equal.

\end{document}